\documentclass[useAMS,usenatbib,usegraphicx]{mn2e}

\newcommand{\feoh}{\textrm{[Fe/H]}}
\newcommand{\cfe}{{\rm [C/Fe]}}

\newcommand{\afe}{\textrm{[$\alpha$/Fe]}}

\newcommand{\msun}{\ensuremath{M_{\odot}}}

\newcommand{\sit}{{\it s}}

\title[Transition of the Stellar Initial Mass Function]
      {Transition of the Stellar Initial Mass Function Explored with Binary Population Synthesis}
\author[T. Suda et al.]{Takuma Suda$^{1}$\thanks{E-mail:takuma.suda@nao.ac.jp},
        Yutaka Komiya$^{1}$,
		Shimako Yamada$^{2}$,
        Yutaka Katsuta$^{2}$,
\newauthor 
		Wako Aoki$^{1}$,
        Pilar Gil-Pons$^{3}$,
        Carolyn L. Doherty$^{4}$,
        Simon W. Campbell$^{4}$,
\newauthor
		Peter R. Wood$^{5}$, and
		Masayuki Y. Fujimoto$^{6,7}\thanks{visiting researcher}$\\
$^{1}$National Astronomical Observatory of Japan, Osawa 2-21-1, Mitaka, Tokyo 181-8588, Japan\\
$^{2}$Department of Cosmoscience, Hokkaido University, Kita 10 Nishi 8, Kita-ku, Sapporo 060-0810, Japan\\
$^{3}$Universitat Politecnica de Catalunya, Campus Baix Llobregat, Building C3, 08860 Castellefels, Spain\\
$^{4}$Monash Centre for Astrophysics (MoCA), Monash University, Victoria 3800, Australia\\
$^{5}$Australian National University, Cotter Road, Weston Creek ACT 2611, Australia \\
$^{6}$Nuclear Data Center, Hokkaido University, Kita 10 Nishi 8, Kita-ku, Sapporo 060-0810, Japan \\
$^{7}$Faculty of Engineering, Hokkaigakuen University, 4-1-40, Asahimachi, Toyohira-ku, Sapporo, 062-8605, Japan
}
\begin{document}

\date{Accepted 1988 December 15. Received 1988 December 14; in original form 1988 October 11
; draft version of 2013 February 11}

\pagerange{\pageref{firstpage}--\pageref{lastpage}} \pubyear{2002}

\maketitle

\label{firstpage}

\begin{abstract}

The stellar initial mass function (IMF) plays a crucial role in
determining the number of surviving stars in galaxies, the chemical
composition of the interstellar medium,
and the distribution of light in galaxies.  A key unsolved question is whether the
IMF is universal in time and space.  Here we use state-of-the-art
results of stellar evolution to show that the IMF of our Galaxy made a
transition from an IMF dominated by massive stars to the present-day
IMF at an early phase of the Galaxy formation.  Updated results from
stellar evolution in a wide range of metallicities have been
implemented in a binary population synthesis code, and compared with
the observations of carbon-enhanced metal-poor (CEMP) stars in our
Galaxy.  We find that applying the present-day IMF to Galactic halo
stars causes serious contradictions with four observable quantities
connected with the evolution of AGB stars.  Furthermore, a
comparison between our calculations and the observations of CEMP stars
may help us to constrain the transition metallicity for the IMF which
we tentatively set at $\feoh \approx -2$.  A novelty of the current
study is the inclusion of mass loss suppression in intermediate-mass
AGB stars at low-metallicity.  This significantly reduces the
overproduction of nitrogen-enhanced stars that was a major problem in
using the high-mass star dominated IMF in previous studies.  Our results
also demonstrate that the use of the present day IMF for all time in
chemical evolution models results in the overproduction of Type I.5
supernovae.  More data on stellar abundances will help to understand
how the IMF has changed and what caused such a transition.

\end{abstract}


\section{Introduction}

The IMF is one of the most important factors influencing the evolution
of galaxies.  In particular, possible transitions of the stellar IMF
from that of the first (metal-free) stars to that of the present-day (metal-rich)
stars have been given much attention.  The typical first
generation stars are thought to be predominantly of
high-mass \citep[e.g. $\sim 40 \msun$ according to][]{Hosokawa2011}
from studies of star
formation in metal-free environments.  However, the actual form of the
IMF for such stars is poorly understood and is still controversial
\citep[see e.g.][and references therein]{Greif2012}.  In addition,
there are few studies dealing with the transition of the IMF during
the early epoch of the Galaxy.  If the IMF is not universal then a
question arises - how and when did the IMF change to the present-day
IMF which is biased towards low-mass stars ($< 0.8 \msun$)
\citep{Salpeter1955}?

Extremely metal-poor stars (EMP stars) observed in the Galaxy can help
to answer this question because we can estimate the number of former
AGB stars that contributed to the change of the surface abundances
evident in these stars.  Spectroscopy of EMP stars \citep{Beers2005a}
reveals that carbon-enhanced stars are very common amongst EMP stars
(20 \% or more \citep{Suda2011,Carollo2012}) when compared to the
higher-metallicity subgiant CH stars ($\sim 1$ \%) \citep{Luck1991}.
The large frequency of carbon-enhanced metal-poor (CEMP) stars is
argued to be a consequence of binary mass transfer from AGB stars that
have undergone carbon dredge-up into their envelopes
\citep{Fujimoto2000,Suda2004,Lucatello2005a,Komiya2007}.  The mechanism for carbon enhancement,
the third dredge-up (TDU), is the currently accepted theory that
explains the origin of classical Ba II stars and CH stars
\citep[see, e.g.][]{McClure1990}.  In the same manner, the origins of some CEMP
stars are thought to be explained by the binary scenario, i.e., the
observed stars have binary companions that are now unseen white
dwarfs.  One of the advantages of this scenario is that we do not
expect all the observed EMP stars to show carbon enhancement on their
surface.  This is qualitatively consistent with the recent discovery
of EMP stars in the lowest metallicity range without large
enhancements of carbon \citep{Caffau2011}.  Binarity has been
statistically confirmed for the population of CEMP stars that show
enhancements of \sit-process elements \citep{Suda2004,Lucatello2005a}, as is
expected from theory.  These objects are called CEMP-\sit\ stars,
in contrast to the other subclass of CEMP stars, the CEMP-no
stars, which do not show any enhancement of \sit-process elements.
The origin of CEMP-no stars is still controversial \citep[see,
  e.g.,][]{Norris2013a}.  Observationally, there is another class of
EMP stars that show enhancements of nitrogen greater than the
enhancement of carbon.  These stars are called nitrogen-enhanced
metal-poor (NEMP) stars and are thought to be produced via binary mass
transfer from intermediate-mass AGB stars which have converted C to N
via hot bottom burning (HBB) during the AGB phase.

Previous works exploring the possibility that the IMF was different in
the early universe \citep{Lucatello2005b,Komiya2007} dealt only with
the statistics of CEMP stars.
(\citet{Komiya2009a} also pointed out that the different IMF is
suggested by the number of low-mass
star survivors estimated from the survey areas of HK and Hamburg-ESO surveys.)
In particular, the IMF which was
proposed to have a peak at around $10 \msun$ \citep{Komiya2007} has
been criticized due to the possible overproduction of NEMP stars
\citep{Izzard2009,Pols2012}.  However recent work on pulsation driven
mass loss in AGB stars at metallicities below $\feoh \sim -2.5$
\citep{Wood2011} suggests that NEMP star formation should have been
suppressed in the early Universe.  That study showed that the growth
of pulsation during the AGB phase is too small to trigger dust-driven
mass loss in low-metallicity, intermediate-mass stars.  With mass-loss
suppressed, the degenerate cores of these stars would have grown to be
massive enough for the stars to explode as type I.5 supernovae, i.e.,
the mass of their cores would have grown to approach the Chandrasekhar
mass limit \citep{Arnett1969,Iben1983a}.  The purpose of this study is
to explore the possibility that the IMF is different at
low-metallicity by including as many ingredients related to stellar
evolution and binary evolution as possible.  In particular, this is
the first study to include characteristic AGB yields at $\feoh < -2.5$
\citep[see, e.g.,][]{Fujimoto2000}.

\section{Binary population synthesis of EMP stars: Code Details}\label{sec:model}

We use the Monte Carlo method to simulate the evolution of binary systems by generating binaries with a range of initial primary mass, secondary mass, and binary period.
The mass of the primary is described by a log-normal function or power-law function.
The former is defined by $dN /d \log m \propto \exp(- (\log(m / \mu))^{2} / (2 \sigma^{2}))$.
The median mass and the dispersion are set at ($\mu$ (in $\msun$), $\sigma$) = (5, 0.6), (10, 0.4), (20, 0.45), (30, 0.5), and (50, 0.6), respectively, in order to reproduce CEMP-no/CEMP-\sit\ ratio following \citet{Komiya2007}.
Power-law functions are defined by $dN / d \log m \propto m^{-x}$, where we choose $x = 1.35$, 0.85, 0.35, and 0.0.
The lower and upper cut-off mass are set at $0.08 \msun$ and $200 \msun$, respectively.
The mass of the secondary is subject to the mass ratio function which was taken from the \citet{Duquennoy1991} or \citet{Raghavan2010}.
For the \citet{Raghavan2010} distribution we adopted a flat distribution by ignoring the preference for like-mass pairs.
For the period distribution of binaries, we adopted a fitting formula for nearby main-sequence stars \citep{Duquennoy1991,Raghavan2010,Rastegaev2010}.

The simulations generate binary systems for metallicities in the
range $-6 \leq \feoh \leq -1$ in steps of 0.5 dex in each simulation.
The total number of binary systems for a given metallicity is typically
one million, which produces more than 1,000 observable giants.  We
note here that
the simulations are compared to observations of giant
stars (Figure~\ref{fig:cfreq}).  Giants were chosen so that the
uncertainty concerning the depth of the convective envelopes for
main-sequence stars is reduced,
and because the uncertainty in the elemental abundances derived for
dwarfs is considerable.  We assume that the stars with $M \geq 0.83 \msun$ evolve to
the final stage of their evolution.

In our simulations, we consider, for the first time, another channel of CEMP star formation, i.e., hydrogen ingestion into the helium flash convective zone for AGB stars at $\feoh \leq -2.5$ (He-Flash Driven Deep Mixing; He-FDDM) \citep{Fujimoto1990,Fujimoto2000}.
Previous studies have not taken this channel into account \citep{Izzard2009,Pols2012}.
Surface chemical abundances of primary stars are taken from stellar models after the He-FDDM at the beginning of the thermally pulsing AGB (TP-AGB) phase \citep{Suda2010}, or after the TDU during the TP-AGB phase.

The surface abundance evolution for all the stars was taken from the
literature \citep{Suda2010,Karakas2010,GilPons2010,GilPons2013}.  For
models undergoing the He-FDDM event, we took the chemical abundances
just after the dredge-up since this event dominates the change of the
surface composition.  For the TDU event, we took the average
abundances of chemical yields provided by \citet{Karakas2010}.  For
models with $\feoh < -2.5$ with the mass range responsible for the
TDU, we adopted the same yields as in \citet{Karakas2010} by adopting
the values at $\feoh = -2.3$ of the same mass.  We also considered the
effect of carbon enhancement by super-AGB stars that host
oxygen-neon-magnesium cores
\citep{Ritossa1999,GilPons2010,GilPons2013}, although it turned out to
be a minor contribution to CEMP stars in our simulations.  A linear
interpolation is made for the surface abundances between the model
grids of initial mass and metallicity.

Binary mass transfer events are treated as instantaneous episodes.
They are described by the masses of the binary components and the
binary separation assuming Kepler rotation without any orbital
ellipticity.  If the binary separation is larger than the sum of the
Roche radius of the primary star and the radius of the secondary star
during the main sequence phase, wind accretion is assumed to be at
play.  Otherwise, Roche lobe overflow (RLOF) applies to the system.
In this study, the occurrence of mass loss in AGB stars is controlled
by a parameter introduced to take into account of the possible
suppression of mass loss at low metallicity \citep{Wood2011}.  Here we
assume that stars with $5 < M / \msun < 8$ and $\feoh \leq -2.5$ do
not eject any envelope and explode as Type I.5 supernovae, which is
chosen to be consistent with the measured abundances using the SAGA
database \citep{Suda2008}.
The SAGA database includes 451 giant stars with carbon abundances
derived from spectra with a resolution of $R = \lambda / \Delta \lambda > 20000$.
The database covers the data published up to 2011 (August, 2012 update).

For wind accretion episodes we assume that the mass accretion onto the secondary star is subject to Bondi-Hoyle accretion with the constant stellar wind velocity of 20 km/s from the primary star.
The amount of mass lost from the primary is estimated from initial-mass final-mass relations for white dwarfs \citep{Han1994}.
These treatments are the same as in our previous work \citep{Komiya2007} and the details are given in the reference \citep{Komiya2007,Izzard2009,Eggleton1983,Hurley2002}.

For RLOF, the Roche radius is calculated from a commonly used approximation , $R_{1} / a = 0.49 q^{2/3} / (0.6 q^{2/3} + \ln (1 + q^{1/3}) )$, where $R_{1}$, $a$, and $q = M_{1} / M_{2}$ are the radius of a primary, binary separation, and mass ratio, respectively, \citep{Eggleton1983}.
If the condition for RLOF accretion is satisfied, we assume a constant mass accretion of $0.05 \msun$ onto the secondary occurs, following \citet{Izzard2009}.
There is however no supporting evidence for this exact amount.
An important point here is that the common envelope phase can be avoided only for binaries that have a mass ratio very close to unity.
Thus the effect of RLOF accretion is actually always negligible in this study, and most binaries undergoing RLOF are excluded from the observable population due to merging.

The definition of CEMP giants is given by $\cfe \geq 0.7$
\citep[see also Aoki et al. 2007]{Suda2011} in the mass range 0.82 to $0.83 \msun$,
considering the stellar lifetimes in the Galactic halo
population (comparable to the age of the Universe).
Self-pollution by AGB stars is excluded due to their negligible contribution to the
population compared with red giants.  We designate CEMP star candidates
as NEMP stars if the initial mass is in the range in which HBB
operates.  This mass range is a parameter of our model and is set at
$4.5 < M / \msun < 8$ in our fiducial models.  For the definition of
CEMP-\sit\ and CEMP-no stars, we use parameters for mass and
metallicity.  We tentatively define a critical mass, set at $3.5
\msun$, below (above) which the model stars become CEMP-\sit\ (no)
stars during the AGB phase.  For CEMP-no stars, it is required that
the model stars should have $\feoh \leq -2.5$ to be consistent with
observations.  Note that the origin of CEMP-no stars is still
controversial \citep[see, e.g.][]{Norris2013a} and will be discussed
in a separate paper.  See also discussions in \citet{Suda2004} for
more detail.

\section{Results and discussion}\label{sec:results}

\begin{figure*}
  \begin{center}
    \includegraphics[width=0.8\textwidth]{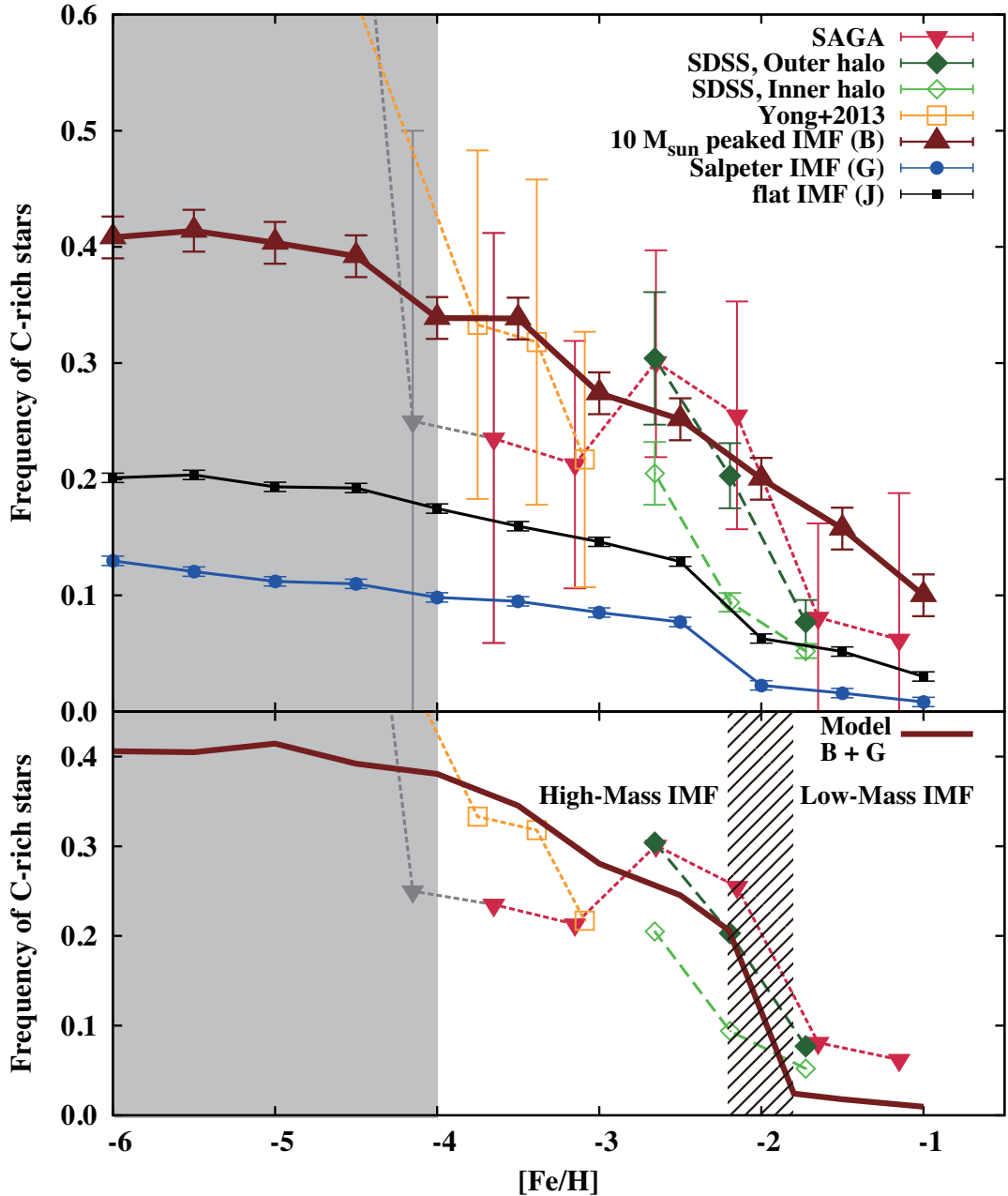}
  \end{center}
  \caption{
Top panel: the CEMP fraction CEMP/EMP for models with
a log-normal IMF with the peak mass at $10 \msun$ (filled triangles,
red), a flat IMF (filled squares, black), and a Salpeter IMF (filled
circles, blue).  The model names in the top-right corner
correspond to those in Table~\ref{tab:model}.
The observed CEMP
fraction from the SAGA database \citep{Suda2008} is shown by filled inverted triangles (red) 
and the observed CEMP fraction from \citet{Yong2013a} is shown by open squares
(orange).  The CEMP fractions for the
outer halo and the inner halo, kinematically selected from the SDSS sample
\citep{Carollo2012}, are shown by filled and open diamonds,
respectively (both green).  The error bars for models and the SAGA
sample are based on the bootstrap method with 1000 trials.
For $\feoh \leq -4$ (shaded area), the number of stars in
the observed sample is less than 10 in each bin and statistically
unreliable.  The criterion for CEMP stars is set at $\cfe \geq 0.7$
\citep{Suda2011} for both models and observations.  Bottom panel:
the CEMP fractions for a model where the transition of the IMF at $\feoh = -2$ is taken into account by
combining the results for the high-mass IMF (model B) and the low-mass
IMF (model G).  The combined model is compared with the same set of
observations as in the top panel.  The hatched area around $\feoh = -2$ represents the
possible transition metallicity of the IMF from the high-mass star
dominated IMF to the present-day IMF.
  }
  \label{fig:cfreq}
\end{figure*}

\begin{figure*}
  \begin{center}
    \includegraphics[width=0.8\textwidth]{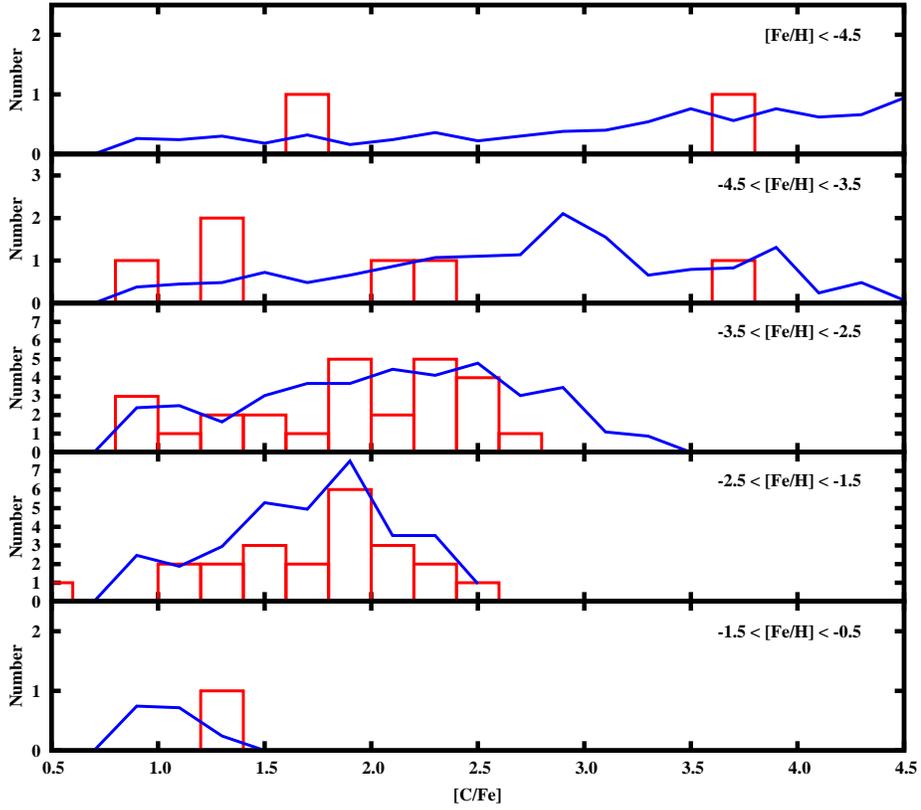}
  \end{center}
  \caption{
		  The comparison of the observed carbon abundance distribution and the distribution of carbon abundance for the fiducial model B.
		  Histograms show the observed distributions for CEMP giants taken from the SAGA database, as used in Fig.~\ref{fig:cfreq}.
		  Model results are shown by solid lines where the number of stars is arbitrarily scaled.
  }
  \label{fig:distc}
\end{figure*}

The frequency of CEMP stars as a function of metallicity is given in
Figure~\ref{fig:cfreq} for observations and for our models with different
adopted IMFs.
The CEMP fraction CEMP/EMP and the fractions CEMP-no/CEMP
and NEMP/CEMP
from our models are shown in Table~\ref{tab:model} for a selected set
of parameters. These ratios are averaged values from the SAGA database
for metallicities below $\feoh = -2$. They are obtained by taking a weighted average
of the ratios in each of the metallicity bins below $\feoh = -2$,
with the weight being the number of stars in each bin.
Model B is our fiducial model as it satisfactorily
reproduces the observations.
It was chosen to reproduce the distribution of carbon abundance as shown in
Figure~\ref{fig:distc}.  It is to be noted that the CEMP
fractions of stars taken from the SAGA database can be affected by selection
bias, although they are consistent with the values derived for other
homogeneous datasets \citep{Yong2013a,Carollo2012}.

The results presented in the top panel of Fig.~\ref{fig:cfreq} demonstrate
that the CEMP fraction for EMP stars is well reproduced by the
high-mass star dominated IMF (model B), while the Salpeter IMF (model G) produces a very low
CEMP fraction compared with observations.  This is consistent with
previous work \citep{Izzard2009,Pols2012}, although we note that those studies
did not consider the effects of stellar evolution at $\feoh <
-2.5$ and assumed that all the stars were born as binaries.  The CEMP
fraction produced by the model with the flat IMF (model J) is also below the observed value for
$\feoh \leq -2$.  These results show that in order to reproduce the CEMP fraction at
$\feoh \leq -2$, the EMP population
should be dominated by massive stars that end their lives as type II
supernovae i.e. a high-mass star dominated IMF is required.
For more metal-rich populations ($\feoh > -2$) such as the inner halo population,
a flat IMF or low-mass star dominated IMF gives a better agreement
with observations.  This implies that an IMF transition occurred at
$\feoh \sim -2$, perhaps in association with the structure formation process
of our Galaxy \citep[see also][]{Yamada2013}.  In the bottom panel of
Fig.~\ref{fig:cfreq}, we present a model with a transition of the
IMF at $\feoh = -2$.

Here we briefly mention the parameter dependencies of the simulations,
although these are thoroughly discussed in a separate paper (Suda et
al., in prep.).  For the choice of the period distribution function,
the distribution derived by \citet{Rastegaev2010} gives preference to
short period binaries, which results in efficient channels for the
common envelope phase and CEMP star formation.  We adopted this
distribution because it is the only study
that considers a population of metal-poor binary stars.  Using the other
two binary period distributions \citep{Duquennoy1991,Raghavan2010} will decrease the fraction of CEMP
stars by up to $\sim$10\% for the high-mass IMF and $\sim$3\% for
the low-mass IMF.  Our fiducial model adopts other input physics and
parameters so that the number of CEMP stars can be maximized with
reasonable choices.  The choice of the mass ratio function does not
significantly alter the fraction of CEMP stars as
discussed in \citet{Komiya2009a} and \citet{Pols2012}.  The boundary
masses for the occurrence of HBB and suppression of mass loss
significantly affect the fraction of NEMP stars.  These values are
still open questions and cannot be determined at this stage
\citep[see][]{Pols2012,Wood2011}.  The mass boundary for the
progenitors of CEMP-no stars is also an open question and can change
the ratio CEMP-no/CEMP.

There are four observable constraints on the simulations.  Three of
them are shown in Table~\ref{tab:model}, and the other one is the
\afe\ abundances of EMP stars.  Firstly, it can be seen that the
CEMP fraction in our models can never reach as high as the
observed 25 per cent when using a low-mass IMF (see also
Fig.~\ref{fig:cfreq}).  In order to check the robustness of the
conclusion, we have tested extreme assumptions for the model
parameters, which include (1) a binary rate of 100 \%, (2) excluding
the contribution of supernova binaries because the speed of companion stars
could have been higher than the velocity dispersion of the host halo
after the disruption of binaries
\citep{Tauris1998}, and (3) the carbon abundance of primary star
enhanced by a factor of 10.  The superposition of all these
assumptions still fails to explain the observed CEMP frequencies at
low-metallicity when a low-mass IMF is used.  A significant increase in the CEMP fraction can only be
obtained by assuming that all the EMP stars are born as binary
systems, although this gives the CEMP fraction well below 20 \% at
$\feoh = -2.5$.  A 100\% binary fraction is however highly unlikely
\citep{Duquennoy1991,Raghavan2010,Rastegaev2010}.

Secondly, we find that the ratio CEMP-no/CEMP is extremely low for
the models using the low-mass IMF.  To obtain a ratio CEMP-no/CEMP
$= 0.5$, we have to set the lower mass boundary for the formation of
CEMP-no stars as small as $\sim 1 \msun$.  However, this then leads to
a mismatch with the observed ratio CEMP-\sit\ /EMP.  Thus it
is evident that other sources of CEMP-no stars, for example stars that
formed out of an ISM pre-enriched with carbon, are needed if we adopt
a low-mass IMF \citep{Meynet2010}.  We note however that any source of
pre-enrichment would require more massive stars, i.e., a massive-star
dominated IMF, although its shape depends on the efficiency of low-mass star
formation out of ejecta from the massive stars.

Thirdly, the ratios of NEMP to CEMP stars in models with low-mass IMFs are too small to be compatible
with the observations.  The observed ratio is around $\sim 0.1$ in the
current available data \citep{Suda2011,Pols2012}, while it was previously thought
to be much smaller \citep{Johnson2007}.  The mass range of NEMP star
progenitors would need to be $2.5 \leq M / \msun \leq 8$, without the assumption of mass loss suppression, to obtain a
NEMP/CEMP ratio $\sim$0.1.
This mass range would need a higher efficiency of HBB at low
metallicity as has been reported by some studies \citep{Campbell2008}.
However, this range will also reduce the CEMP fraction dramatically, and
hence cannot coexist with the requirement for the large frequency of
CEMP stars. 

Finally, the number of the progenitors of type I.5 supernovae
dominates over, or is comparable to, that of type II supernovae when a low-mass IMF is used, which
is not supported observationally since Galactic halo stars do not show the
low values of \afe\ that would be expected from type I.5 supernovae
\citep{Nomoto1984}.
Thus our simulations arrive at the robust conclusion that
population simulations using low-mass IMFs cannot match the
observational constraints given by Galactic EMP stars.

Our results add to previous evidence for a non-universality of
the IMF.  Other evidence comes from studies such as the population of
white dwarfs \citep{Adams1996}, those involving simulations of star
formation \citep[see, e.g.][]{Larson2005}, and observations of
early-type galaxies \citep[see, e.g.][]{vanDokkum2008}.  Our study
shows that the observed population in the Galactic halo can be
explained by a transition of the IMF from one biased towards high
masses to one biased toward lower masses at a metallicity of $\feoh
\sim -2$.  We propose that the statistics of CEMP and NEMP stars in
the entire range of metallicity in our Galaxy and dwarf galaxies in
the local group should be investigated to reveal the star formation
history of galaxies.

\section{Conclusions}

We have compared the results of our binary population synthesis
model with the observations of metal-poor halo stars in the Galaxy.  We find
that a low-mass biased IMF, similar to the present one, is
incompatible with the four observable quantities related to the CEMP
subclasses.  On the other hand, considering a
high-mass biased IMF, together with the occurrence of SN I.5 in a
certain mass and metallicity range (supported by recent simulations of
pulsation-driven mass-loss in low metallicity stars), a more
reasonable agreement between models and observations is found.  Our
results imply that the transition of the IMF occurred around a
metallicity of $\feoh \sim -2$, which should be tested by star
formation theory and observations.  These results also give important
clues in the most metal poor
regimes as to the mass and metallicity limits of stars that
experience HBB, TDU and stellar winds, or that allow the formation of
CEMP (-\sit\ and -no) stars,  NEMP stars,  SN I.5 and SNII.
Our understanding of these limits has been traditionally hampered by the
scarcity of observations.  We hope these results can help to improve
the understanding of the chemical evolution of our Galaxy.

\section*{Acknowledgments}

The authors thank Daniela Carollo and David Yong for kindly providing the data of CEMP fraction.
We also thank Takayuki Saitoh for useful comments and suggestions.
This work has been supported by Grant-in-Aid for Scientific Research (18104003), from Japan Society of the Promotion of Science.
T.S. acknowledges the support of the Institutional Program for Young Researchers Overseas Visits by JSPS, which enhanced the results and discussions in this work.

 \newcommand{\noop}[1]{}

\clearpage

\begin{table}
    \begin{minipage}{70mm}
    \caption{Characteristics of simulated EMP stars with the different IMF}
    \label{tab:model}
    \begin{tabular}{lllll}
    \hline
       ID & ($\mu$, $\sigma$) or $x$ & $\frac{\textrm{CEMP}}{\textrm{EMP}}^{d}$ & $\frac{\textrm{CEMP-no}}{\textrm{CEMP}}^{d}$ & $\frac{\textrm{NEMP}}{\textrm{CEMP}}^{d}$ \\
    \hline
       A  &  5, 0.6    & 0.19  & 0.10  & 0.036  \\
       B$^{a}$  & 10, 0.4    & 0.28  & 0.24  & 0.089  \\
       C  & 20, 0.45   & 0.22  & 0.29  & 0.092  \\
       D  & 30, 0.5    & 0.19  & 0.29  & 0.104  \\
       E  & 50, 0.6    & 0.19  & 0.28  & 0.082  \\
       F$^{b}$  & 0.79, 0.51 & 0.037 & 0.00  & 0.072  \\
    \hline
       G$^{c}$  & 1.35       & 0.08  & 0.02  & 0.010  \\
       H  & 0.85       & 0.10  & 0.02  & 0.013  \\
       I  & 0.35       & 0.12  & 0.04  & 0.016  \\
       J  & 0.0        & 0.14  & 0.07  & 0.032  \\
    \hline
       Observed &      & 0.25  & 0.5   & $\sim$0.1 \\
    \hline
   \end{tabular}
   \medskip
    $^{a}$ \citet{Komiya2007} \\
    $^{b}$ \citet{Lucatello2005b} \\
    $^{c}$ \citet{Salpeter1955} \\
    $^{d}$ Averaged values considering the metallicity distribution function of the metal-poor stars for $\feoh \leq -2$. See text.
   \end{minipage}
\end{table}

\clearpage

\clearpage

\end{document}